\documentclass{article} 
\usepackage{float}
\usepackage[preprint]{colm2025_conference}
\usepackage{multirow}
\usepackage{wrapfig}
\usepackage{microtype}

\usepackage{hyperref}
\usepackage{url}
\usepackage{booktabs}
\usepackage{amsmath}
\usepackage{lineno}
\usepackage{tgheros}
\usepackage{algorithm}
\usepackage{booktabs}
\usepackage{algpseudocode}
\usepackage{xcolor}
\usepackage[most]{tcolorbox}
\usepackage{tikz}
\usepackage{wrapfig}
\usepackage{lipsum} 
\usetikzlibrary{positioning, arrows.meta, quotes}

\usepackage{tikz} 

 \newcommand{\squishlist}{
	\begin{list}{$\bullet$}
		{ \setlength{\itemsep}{0pt}
			\setlength{\parsep}{3pt}
			\setlength{\topsep}{3pt}
			\setlength{\partopsep}{0pt}
			\setlength{\leftmargin}{1.5em}
			\setlength{\labelwidth}{1em}
			\setlength{\labelsep}{0.5em} } }
	\newcommand{\squishlisttwo}{
		\begin{list}{$\bullet$}
			{ \setlength{\itemsep}{0pt}
				\setlength{\parsep}{0pt}
				\setlength{\topsep}{0pt}
				\setlength{\partopsep}{0pt}
				\setlength{\leftmargin}{2em}
				\setlength{\labelwidth}{1.5em}
				\setlength{\labelsep}{0.5em} } }
		\newcommand{\squishend}{
	\end{list}  }

\definecolor{darkblue}{rgb}{0, 0, 0.5}
\hypersetup{colorlinks=true, citecolor=darkblue, linkcolor=darkblue, urlcolor=darkblue}

\title{Accelerating Causal Network Discovery of Alzheimer’s Disease Biomarkers via Scientific Literature-based Retrieval Augmented Generation}

\author{
Xiaofan Zhou \\
University of Illinois Chicago \\
1200 W Harrison St, Chicago, IL 60607 \\
\texttt{xzhou77@uic.edu}
\And
Liangjie Huang \\
University of Illinois Chicago \\
1200 W Harrison St, Chicago, IL 60607 \\
\texttt{lhuan85@uic.edu}
\And
Pinyang Chen \\
Pennsylvania State University \\
201 Old Main, University Park, PA 16802 \\
\texttt{ppc5322@psu.edu}
\And
Wenpeng Yin \\
Pennsylvania State University \\
201 Old Main, University Park, PA 16802 \\
\texttt{wenpeng@psu.edu}
\And
Rui Zhang \\
Pennsylvania State University \\
201 Old Main, University Park, PA 16802 \\
\texttt{rmz5227@psu.edu}
\And
Wenrui Hao \\
Pennsylvania State University \\
201 Old Main, University Park, PA 16802 \\
\texttt{wxh64@psu.edu}
\And
Lu Cheng \\
University of Illinois Chicago \\
1200 W Harrison St, Chicago, IL 60607 \\
\texttt{lucheng@uic.edu}
}

\begin{document}

\ifcolmsubmission
\linenumbers
\fi

\maketitle

\begin{abstract}
The causal relationships between biomarkers are essential for disease diagnosis and medical treatment planning. One notable application is Alzheimer's disease (AD) diagnosis, where certain biomarkers may influence the presence of others, enabling early detection, precise disease staging, targeted treatments, and improved monitoring of disease progression. However, understanding these causal relationships is complex and requires extensive research. Constructing a comprehensive causal network of biomarkers demands significant effort from human experts, who must analyze a vast number of research papers, and have bias in understanding diseases' biomarkers and their relation.  
This raises an important question: Can advanced large language models (LLMs), such as those utilizing retrieval-augmented generation (RAG), assist in building causal networks of biomarkers for further medical analysis? To explore this, we collected 200 AD-related research papers published over the past 25 years and then integrated scientific literature with RAG to extract AD biomarkers and generate causal relations among them. 
Given the high-risk nature of the medical diagnosis, we applied uncertainty estimation to assess the reliability of the generated causal edges and examined the faithfulness and scientificness of LLM reasoning using both automatic and human evaluation.
We find that RAG enhances the ability of LLMs to generate more accurate causal networks from scientific papers. However, the overall performance of LLMs in identifying causal relations of AD biomarkers is still limited. We hope this study will inspire further foundational research on AI-driven analysis of AD biomarkers causal network discovery.
\end{abstract}

\section{Introduction}
Alzheimer's disease (AD) is a complex neurodegenerative disorder characterized by intricate interactions among various biomarkers \citep{wang2023biomarkers,hao2016mathematical}. Understanding these interactions is crucial for early diagnosis, disease progression modeling, and therapeutic development \citep{zheng2022data,petrella2019computational}. Causal network construction provides a powerful framework for uncovering these relationships, yet existing approaches \citep{spirtes1991algorithm,strobl2018fast,chickering2020statistically} often rely solely on structured data, overlooking the vast knowledge embedded in scientific literature.  

This limited understanding of AD has become a significant obstacle in devising effective therapeutic strategies to impede its progression. Despite publishing over 5,000 AD research articles annually \citep{wu2023examination}, the integration of AD pathophysiology remains incomplete. Trained from extensive observations of the world, large language models (LLMs) exhibit remarkable abilities in understanding unstructured data and utilizing the learned knowledge to address various broad tasks \citep{bubeck2023sparks}. In this work, we pioneer the use of retrieval-augmented generation (RAG) \citep{lewis2020retrieval,gao2023retrieval} to synthesize information from AD scientific literature, developing a comprehensive causal network of AD biomarkers that summarizes AD pathophysiology. By leveraging scientific literature and domain expertise, we introduced both standard RAG and a streamlined split-RAG frameworks to improve causal network discovery in AD research. Different from the standard RAG, which uses all retrieved documents as context, split-RAG processes each document individually to generate answers, subsequently aggregating these through methods such as majority voting. This approach addresses the challenge smaller LLMs face in understanding long context, providing a simpler and shorter context for more accurate identification. Our approach is driven by interdisciplinary collaboration, ensuring that both data-driven insights and expert knowledge contribute to the network construction process. 

Our work concludes following contributions:
\squishlist

\item We pioneer exploring RAG for accelerating AD biomarkers' causal discovery, integrating literature-based knowledge extraction with LLMs' reasoning capabilities.  

\item We build a dataset of 200 AD-related papers from 2000 to 2025, selecting 40 papers per five-year interval. This dataset serves as a valuable resource for AD biomarker extraction and causal analysis.  

\item We design a framework utilizing RAG to extract AD biomarkers from the collected papers and identify their causal relationships. This framework aids in AD diagnosis analysis and tracks the evolution of knowledge in AD biomarkers.

\item To measure the trustworthiness of the RAG framework, we apply uncertainty estimation to assess the reliability of predicted causal edges as well as automatic and human evaluations to assess the faithfulness and scientificness of LLM's reasoning, respectively.

\squishend 

By integrating RAG with scientific literature and domain expertise, our work bridges the gap between unstructured text-based knowledge and computational causal discovery, offering a scalable and interpretable approach to studying AD biomarkers.  

\section{Related work}
\subsection{LLM for causal discovery}
There are many attempts to apply LLMs to causal discovery. For example, \cite{kiciman2023causal} find that LLMs can generate more accurate causal arguments than conventional methods. \cite{choi2022lmpriors} leverage meta-data (e.g., variable names and descriptions) for causal discovery. \cite{long2023can} integrate causal discovery with large-scale biomedical knowledge graphs to uncover causal relationships among biomarkers with the help of LLM. \cite{jiralerspong2024efficient} propose a more efficient approach for generating causal networks by leveraging statistical information in prompt for LLM while avoiding cycles. In contrast, \cite{zhang2024causal} introduces a pairwise LLM-based causal network construction method that lacks efficiency. \cite{gao-etal-2023-chatgpt} report that LLMs are poor causal reasoners but perform relatively well as causal interpreters, although their output is often affected by hallucinations. \cite{jin2024can} show that LLMs perform close to random on causal inference tasks, and that self-correction does not significantly improve results. \cite{long2022can} evaluate GPT-3’s ability to construct causal relations. The work most relevant to ours is \cite{zhang2024causal} which also employs RAG to build causal networks through pairwise predictions. Different from it, our work pioneers the application of RAG for AD biomarker causal discovery which involves both scientific literature and domain expertise, as well as uncertainty estimation and reasoning analyses to study its trustworthiness.
\subsection{RAG for science}
RAG \citep{lewis2020retrieval} integrates information or knowledge from external data sources, providing supplementary reference or guidance for input queries and generated outputs \citep{khandelwal2019generalization}. RAG-enhanced LLMs have demonstrated their utility in various scientific domains, including biology \citep{wang2024biorag}, chemistry \cite{chen2023chemist}, and geometry \cite{levonian2023retrieval}. For instance, \cite{wang2022retrieval} proposed a retrieval-based framework that leverages database queries to inform and direct molecular generation.
Further, PGRA \citep{guo2023prompt} employs a retriever to search for and re-rank relevant contexts before generating responses. Later research has various approaches, including optimizing retrieval processes by leveraging prior answers \citep{wang2023self}, and improving model functionality via iterative feedback cycles \citep{liu2024ra}. \cite{levonian2023retrieval} investigate the use of RAG to enhance responses to middle-school algebra and geometry questions. \cite{chen2023chemist} automate the recommendation of reaction conditions in chemical synthesis using RAG technology.

We advance prior research by innovatively applying the RAG model to accelerate the construction of causal networks for AD biomarkers, integrating knowledge extraction from literature with LLMs' reasoning capabilities. Our approach includes building a dataset of 200 AD-related papers, developing a RAG-based framework to identify biomarkers and their causal relationships, and bridging the gap between textual knowledge and computational causal analysis, providing a scalable and interpretable method for accelerating causal network discovery of AD biomarkers.
\begin{figure}
    \centering
    \includegraphics[width=1\linewidth]{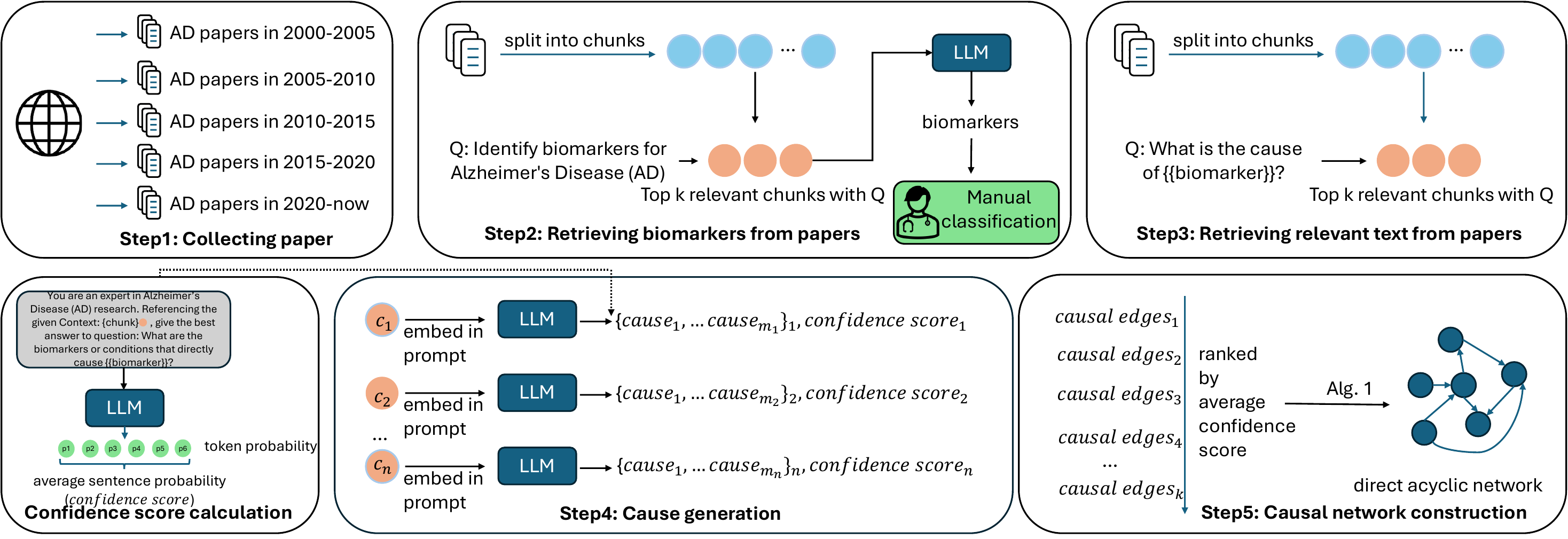}
    \caption{Overview of the proposed framework. $c_1,c_2..,c_n$ are retrieved chunks}
    \label{method}
\end{figure}
\section{Method}
In this section, we detail the framework of leveraging scientific literature-based RAG to accelerate discovering the causal networks of AD biomarkers (Figure \ref{method}). Our framework involves the following major steps: 1) Collecting scientific papers related to AD biomarkers from premier journals. 2) Extracting related biomarkers from these scientific papers. 3) Constructing the causal relations between biomarkers using the collected papers and RAG. 4) Quantifying the uncertainty of each RAG predicted causal edges. 5) Construct a causal network using the causal relationships and associated uncertainty scores derived from the previous step.
\subsection{Scientific papers collection}
Based on citation data, we collected 200 open-access papers on AD biomarkers that are published after 2000 to analyze the evolution of knowledge in the field. The search keywords include \{\textit{Alzheimer disease pathophysiology}, \textit{pathology Alzheimer's disease}, and \textit{pathophysiology of Alzheimer's disease}\}.  The papers were grouped into five five-year intervals: 2000–2005, 2006–2010, 2011–2015, 2016–2020, and 2021–2025, with 40 papers per period. Using Google Scholar citations, we selected the most highly cited papers within each interval for further analysis. These 200 papers serve as our RAG database.

\subsection{RAG method for AD-related information extraction}
In this section, we introduce how to adapt the RAG method in \cite{lewis2020retrieval} for extracting AD-related information from papers. This approach first concatenates text from multiple papers and then splits it into smaller chunks. Each chunk is then projected into an embedding space \( R^d \) using an embedding model, where \( d \) represents the embedding dimension. To retrieve AD-related information, we project the query into the same embedding space using the same model and employ a similarity function (e.g., cosine similarity) to identify the most relevant text chunks. Compared to traditional methods, this approach enhances prediction accuracy, mitigates hallucinations caused by long contexts, and improves knowledge utilization.

For the chunking method, we follow the approach outlined in \cite{sarthi2024raptor}. First, we split the AD paper into chunks by segmenting it into sentences using delimiters such as ".", "!", "?", and "\verb|\n|". If a sentence exceeds a predefined token limit, it is further split using punctuation marks like ",", ";", and ":". In this step, from the given papers, we ultimately obtain a set of chunks \(\{c_1, c_2, c_3, \dots, c_k\}\) which are most likely to contain the relationship among different biomarkers. To project these chunks into embedding space, we utilize a transformer \citep{vaswani2017attention} to obtain their embeddings. 
In our project, we employ a sentence-transformer optimized for semantic search \citep{song2020mpnet}. The chunks achieved in the previous step are projected into embedding space by the transformer method to get a set of embeddings \(\{e_1, e_2, e_3, \dots, e_k\}, e_i\in R^d\).

In our application, we will choose the top-$k$ chunks with the highest similarity score with the question representation. Given a question $q_j$ asking about the relationship between different biomarkers, in this step, the RAG will return a set of retrieved chunks \(\{c_1^j, c_2^j, c_3^j, \dots, c_k^j\}\) that are most related to AD Biomarkers.

\subsection{Biomarkers extraction and causal networks construction}
\label{construct}
For biomarker extraction, we use the following prompt to retrieve relevant chunks from the paper corpus:  

\begin{tcolorbox}[colback=gray!10, colframe=gray!80]
Identify biomarkers for Alzheimer's Disease (AD)
\end{tcolorbox}

Given the broad range of biomarkers extracted from research papers---such as Lobar Atrophy, Hippocampal Atrophy, Microbleed Volume, Synaptotagmin 1, Alpha-Synuclein, Neurogranin (Ng), S100$\beta$ Protein, and Lyn-10 Protein---all of which fall under the \textit{neurodegeneration} category, we consulted with AD clinicians to establish a systematic classification. Based on expert input, we grouped the retrieved biomarkers into six categories: \{\textit{amyloid beta, APOE, tau, neuroinflammation, neurodegeneration}, and \textit{metabolism}\}.  
To study their relationship with AD, we introduce *\textit{cognitive decline and impairment}* as the outcome node, representing the progression of AD.

Given the identified biomarkers, we present our strategy for constructing a causal network using the literature-based RAG. Specifically, for each outcome node in the previous step, we identify potential causal nodes. The following is an example prompt:

\begin{tcolorbox}[colback=gray!10, colframe=gray!80]What are the biomarkers or conditions that directly cause cognition decline? Select from the options in A. amyloid Beta ($\beta$), B.APOE, C.Tau, D.Neuroinflammation, E.cognition decline, F.Neurodegeneration, G.Metabolism.
\end{tcolorbox} 

The above question is then projected into embedding space and used for retrieving the top-$k$ most relevant chunks in the papers corpus. We can then either (1) feed all retrieved chunks into the LLM as the context in the prompt, i.e., \textit{concat-RAG} or (2) generate an output for each retrieved chunk individually and then aggregate the results (e.g., majority vote), i.e., \textit{split-RAG}. We hypothesize that split-RAG is more suitable for smaller LLMs as it would be easier for the LLM to identify causal relations within simpler and shorter context \citep{merth2024superposition}. We also incorporate Chain-of-Thought (CoT) \citep{wei2022chain} to enhance LLM's reasoning performance. The prompt we use for causal discovery for $node_i$ is:


\begin{tcolorbox}[colback=gray!10, colframe=gray!80]
\label{gen}
You are an expert in Alzheimer’s Disease (AD) research. Referencing the given Context: {chunk}, give the best answer to the question: What are the biomarkers or conditions that directly cause {$node_i$}? Select from the options in A. amyloid Beta ($\beta$), B. APOE ... Then, provide your final answer (variable label only) within the tags $<Answer>...</Answer>$, if not find any, return $<Answer></Answer>$. Let’s work this out step-by-step. Your step by step answer is:
\end{tcolorbox} 

\textbf{Uncertainty estimation.} For each node \( node_i \), the LLM will return several biomarkers that can cause \( node_i \), and we compute their average uncertainty score via the averaged token probability of the output sequence following \cite{ott2018analyzing}. Specifically, for a given \( node_i \), each retrieved chunk \( c \) will be fed into the LLM which then generates several cause nodes \( nodes_c \) and an uncertainty score \( u_c \) computed as,
\begin{equation}
\label{uncertain}
u_c = \frac{1}{n} \sum_{i=1}^{n} p_i,
\end{equation}
where $n$ is the number of generated tokens and $p_i$ denotes the probability of token $i$. All the cause nodes produced by this chuck share the same uncertainty score.

A lower average token probability indicates higher uncertainty, while a higher average probability suggests the model is more confident in its predictions.
The final uncertainty score of \( node_j \) causing \( node_i \) is given by
\begin{equation}
uncertaintyscore_{node_j \to node_i} = \frac{\sum_{c \in C_j} u_c}{|C|},
\end{equation}
where \( C \) is the set of retrieved chunks for \( node_i \), and \( C_j = \{c \mid node_j \in nodes_c, c \in C\} \).

Given the above process, we obtain several causal edges and their corresponding uncertainty scores. To build the final causal network, we first rank the causal edges by their uncertainty scores. To ensure that the causal network is a directed acyclic graph (DAG) \citep{neuberg2003causality}, we prioritize adding edges with higher average token probability first. If adding an edge with a lower uncertainty score creates a cycle in the graph, we discard that edge. The pseudo-code of the RAG framework can be found in Algorithm \ref{causal}.

\begin{algorithm}
\begin{algorithmic}[1]
\caption{Causal network discovery for AD biomarkers with RAG}

\label{causal}
\Require RAG model $\rho(\cdot)$ introduced in Sec \ref{construct} in , cycle checker $CheckCycle()$, nodes list $nodeset=\{node_1, node_2,..., node_m\}$
\State$edges \gets \{\}$\Comment{Create an empty edge list to store the result.}
\For{$node$ in $nodeset$}
    \State $causes$, $uncertaintyscores = \rho(node)$
    \For{$cause$, $uncertaintyscore$ in zip($causes$, $uncertaintyscores$)}
        \State $edges.add(cause, node, uncertaintyscore)$
    \EndFor
\EndFor
\State rank $edges$ based on $uncertaintyscore$.
\State $G \gets \{\}$\Comment{Create an empty causal network to store the result.}
\For($edge$ in $edgee$)
    \If{not $CheckCycle(G,edge)$}
        \State$G.add(edge)$
    \EndIf
\EndFor    
\State \Return $G$

\end{algorithmic}
\end{algorithm}

\section{Experiments}

In this section, we present a series of experiments to evaluate the proposed framework on various LLMs, assessing their capability in AD biomarker causal network discovery. We aim to answer the following research questions (RQs):
\squishlist
\item \textbf{RQ1}: How well do concat-RAG and split-RAG perform compared to basic LLMs (base-LLM) for AD biomarkers causal discovery?
\item \textbf{RQ2}: How does the number of retrieved chunks affect causal network performance?
\item \textbf{RQ3}: Is the CoT reasoning faithful and scientific? 
\item \textbf{RQ4}: Based on the results of RAG, how does the causal network of AD biomarkers evolve between 2005-2025? 
\squishend
\subsection{Experimental setup}  

We detail the experimental setup including LLMs used, evaluation metrics, and ground-truth collections in this section.
\paragraph{Models} 
We employ four different LLMs to ensure a balance between efficiency and performance across various tasks. These models are categorized based on their scale:  (1) \textbf{Small-scale models}: Meta-Llama-3-8B \citep{llama3modelcard}: An 8-billion-parameter model from Meta's Llama 3 series, designed for efficiency while maintaining strong reasoning and language understanding capabilities. Mistral-7B-Instruct \citep{jiang2023mistral7b}: A 7-billion-parameter instruction-tuned model by Mistral AI, optimized for following user prompts and handling diverse NLP tasks.  
(2) \textbf{Large-scale models}: Meta-Llama-3.1-70B \citep{grattafiori2024llama}: A 70-billion-parameter model from Meta's latest Llama 3.1 release, providing enhanced contextual understanding, reasoning, and generation quality. Mixtral-8x7B-v0.1 \citep{jiang2024mixtralexperts}: A mixture-of-experts model consisting of eight 7-billion-parameter experts, selectively activating subsets of experts to optimize performance and computational efficiency.  By leveraging models of different sizes, we achieve a trade-off between computational cost and model performance, allowing us to adapt to different scenarios based on task requirements.

\begin{wrapfigure}{R}{0.4\textwidth}
  \centering
  \includegraphics[width=0.38\textwidth]{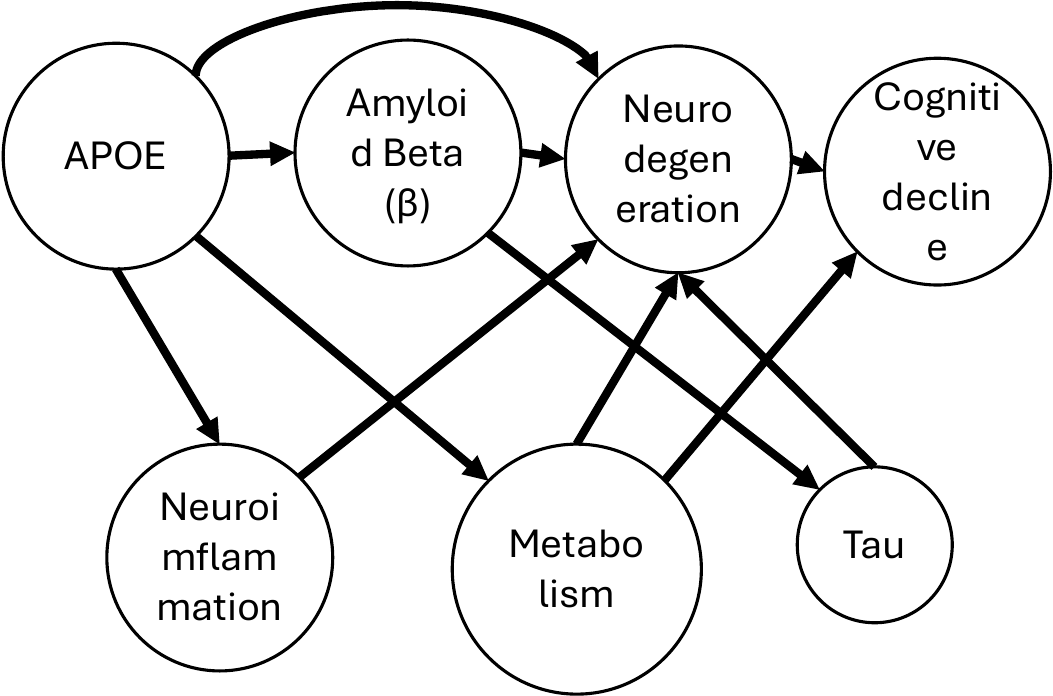}
  \caption{Ground truth of AD biomarker causal network.}
  \label{ground}
\end{wrapfigure}
\paragraph{Ground Truth Collection and Evaluation Metrics} Our evaluation mainly focuses on three aspects: (1) The accuracy of causal network construction. For the accuracy of causal networks, we use the standard metrics for causal discovery including accuracy, F1-score, and precision. To obtain the ground-truth causal network (Figure \ref{ground}), the same AD clinician is asked to construct the causal edges among the six AD biomarkers and the node of cognitive decline and impairment. 
(2) Whether the CoT reasoning is faithful to the retrieved context and whether it is scientific. For the faithfulness of the reasoning, we use the TRUE \citep{honovich-etal-2022-true-evaluating}, a pre-trained T5 model \citep{raffel2020exploring} for assessing factual consistency, to calculate an entailment score following \citep{gao-etal-2023-enabling}. The entailment score can be calculated by:
\begin{equation}
    entailment = \sum_{\text{D}}\frac{1\cdot(TRUE(D_i))}{|D|},
\end{equation}
where \( D = \{\text{retrieved chunks}_i, \text{reasoning}_i\}_{i=1}^{|nodes| \cdot k} \) represents all retrieved chunk and reasoning pairs, with \( k \) denoting the number of top-k chunks retrieved for each node. The same AD clinician is asked to evaluate whether the CoT reasoning is scientific. 
(3) The evaluation of uncertainty estimation. For the evaluation of uncertainty estimation, we use the AUROC metric following~\cite{kaur2024addressing, geng-etal-2024-survey}.
\begin{table}[t!]
    \centering    
    \renewcommand{\arraystretch}{1.2}
    \resizebox{0.7\linewidth}{!}{
    \begin{tabular}{llcccc}
        \toprule
        Model & Approach & Accuracy & Precision & F1-score & AUROC \\
        \midrule
        \multirow{3}{*}{Llama-3.1-70B} 
       
        & base-LLM &0.4600&0.3084&0.3657&0.5309  \\
         & split-RAG & 0.\textbf{6429} & 0.3054 & 0.2937 & 0.3762 \\
        & concat-RAG & 0.3600&\textbf{0.3911}&\textbf{0.3730}&\textbf{0.5653}\\
        \midrule
        \multirow{3}{*}{Mixtral-8x7B-v0.1} 
        
        & base-LLM & 0.1600&0.1650&0.1597&0.3348   \\
        & split-RAG &  \textbf{0.6667} & 0.3227 & \textbf{0.3399} & 0.4636 \\        
        & concat-RAG & 0.2400&\textbf{0.3767}&0.2866&\textbf{0.4779}  \\
        \midrule
        \multirow{3}{*}{Mistral-7B} 
       
        & base-LLM & 0.3600&0.2163&0.2699&0.3830  \\
         & split-RAG &  \textbf{0.6310} & \textbf{0.3179} & \textbf{0.3880} & 0.3736 \\
        & concat-RAG & 0.4400&0.2805&0.3421&\textbf{0.5703} \\
        \midrule
        \multirow{3}{*}{Llama-3-8B} 
        
        & base-LLM & 0.5400&\textbf{0.4481}&\textbf{0.4804}&0.3144 \\
        & split-RAG &  \textbf{0.6429} & 0.3579 & 0.4614 & \textbf{0.4974} \\
        & concat-RAG & 0.3400&0.2471&0.2826&0.4278 \\
        \bottomrule
    \end{tabular}
    }
    \caption{Performance of the constructed causal network across three different strategies and four language models. Bold font indicates the best performance within each LLM model. }
    \label{main1}
\end{table}
\begin{table}[t!]
    \centering
    \renewcommand{\arraystretch}{1.2}
    \resizebox{0.7\linewidth}{!}{
    \begin{tabular}{llcccc}
        \toprule
        Model & top $k$ & Accuracy & F1-Score & Precision & AUROC\\
        \midrule
        \multirow{3}{*}{Llama-3.1-70B} 
        & 10    &  \textbf{0.6429} & 0.3054 & \textbf{0.2937} & 0.3762 \\
        & 20    &  0.5516 & 0.1915 & 0.2153 & 0.3835 \\
        & 50    &  0.5397 & \textbf{0.3390} & 0.2566 & \textbf{0.5016} \\   
        \midrule
        \multirow{3}{*}{Mixtral-8x7B-v0.1} 
        & 10     &  \textbf{0.6667} & 0.3227 & \textbf{0.3399} & 0.4636 \\
        & 20     &  0.5992 & 0.2755 & 0.3350 & \textbf{0.5880} \\
        & 50     &  0.5722 & \textbf{0.3569} & 0.2727 & 0.5426 \\
        \midrule
        \multirow{3}{*}{Mistral-7B} 
        & 10    &  \textbf{0.6310} & 0.3179 & \textbf{0.3880} & 0.3736 \\
        & 20    &  0.5873 & 0.2774 & 0.3469 & 0.4580 \\
        & 50    &  0.5523 & \textbf{0.3996} & 0.2715 & \textbf{0.6194} \\   \midrule
        \multirow{3}{*}{Llama-3-8B} 
        & 10    &  \textbf{0.6429} & 0.3579 & \textbf{0.4614} & 0.4974 \\
        & 20    &  0.6310 & 0.3437 & 0.4404 & \textbf{0.5658} \\
        & 50    &  0.5674 & \textbf{0.4026} & 0.3038 & 0.5244 \\        \bottomrule
    \end{tabular}
    }
    \caption{Impact of the number of retrieved texts with split-RAG. Bold font indicates the best performance within each LLM model.}
    \label{sen}
\end{table}
\subsection{Main results for RAG (RQ1)}
We compare the performance of the RAG methods with base-LLM using different LLMs as the backbone models in Table \ref{main1}. Base-LLM generates the causal networks without using the retrieved context, but its other processes still follow Algorithm \ref{causal}. For both RAG methods and base-LLM, we repeat the generations in Section \ref{gen} 10 times and use the averaged results for every 5 years as the final result. In this experiment, the number of retrieved chunks is 10. From Table \ref{main1}, we have the following findings:

(1) RAG-based methods (split-RAG and concat-RAG) often outperform the base-LLM across all three causal discovery evaluation metrics, except for the Llama-3-8B model. For example, with Llama-70B, split-RAG achieves the best Accuracy and concat-RAG achieves the best Precision and F1 score. Split-RAG also achieves the best performance w.r.t. all three metrics with Mistral-7B. This result demonstrates the advantage of using literature-based RAG to build AD biomarkers causal networks. However, for all three approaches, the causal discovery performance is unsatisfactory, suggesting the limitations of current LLMs in identifying causal relations. 

(2) In most cases, Accuracy in all methods is higher than Precision, suggesting that LLMs is conservative in predicting the positive causal edges (i.e., edges should exist), leading to a high number of true negative (which boosts accuracy) but also high number of false negatives. Split-RAG generally shows a better performance than concat-RAG when LLMs are small in size. We believe this is because aggregating all contexts shall boost LLMs' causal edges extraction performance, but small LLMs are limited in understanding long and possibly contradicting context.

(3) The uncertainty evaluation results of AUROC show that RAG not only predict better than base-LLM but are also capable of indicating when they are unsure about their predictions, suggesting literature-based RAG could enhance the calibration of the model's confidence. Concat-RAG generally outperforms split-RAG in uncertainty estimation.
\begin{figure}
    \centering
    \includegraphics[width=0.9\linewidth]{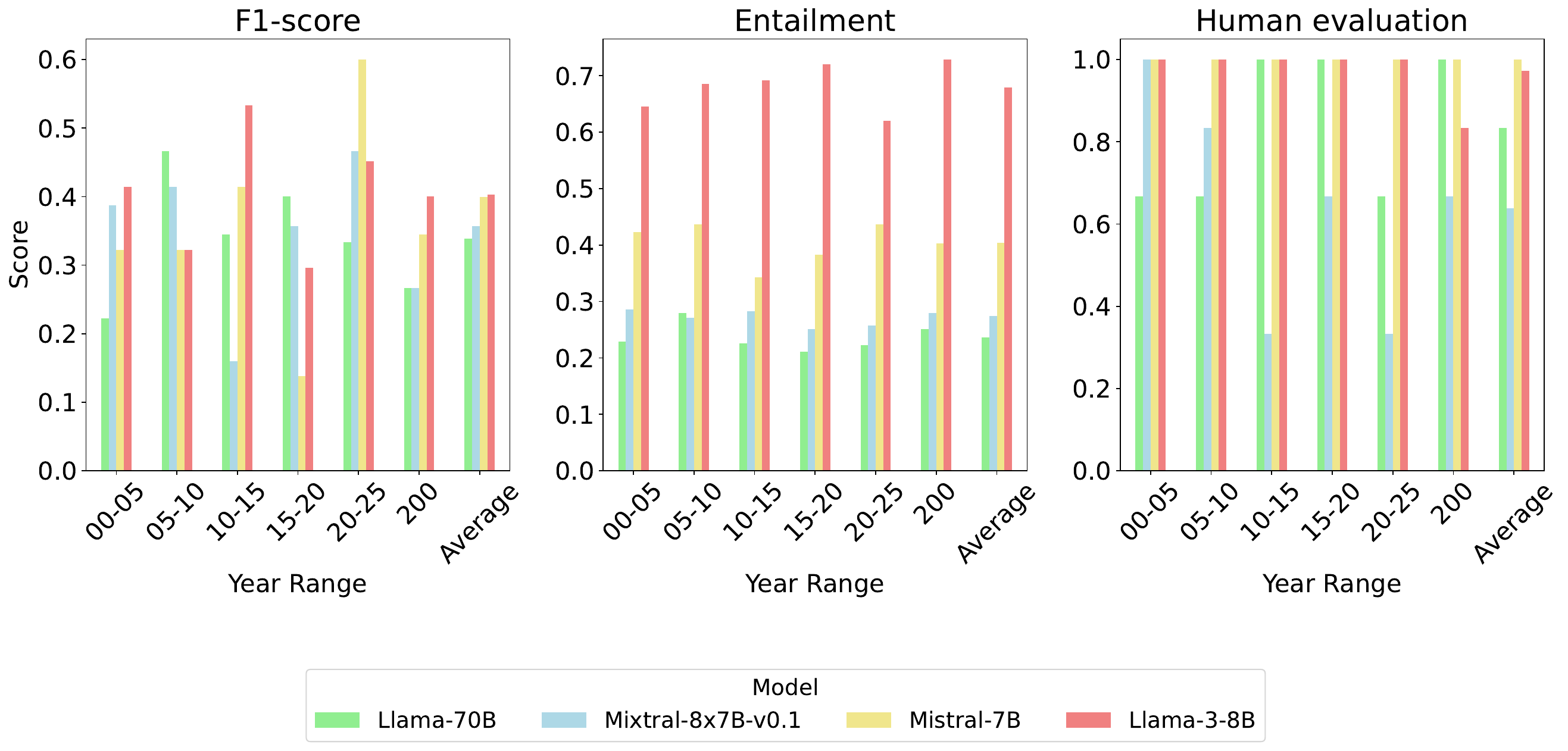}
    \caption{Faithfulness and scientificness analyses of CoT reasoning across different years. }
    \label{main}
\end{figure}
\subsection{Impact of the number of chunks $k$ (RQ2)}
For simplicity, we use split-RAG in all the rest of the experiments.  Here, we investigate the influence of the number of retrieved chunks. We conduct experiments on the four LLMs with $k$ set to \{10, 20, 50\}. The results are shown in Table \ref{sen}. The observation suggests a nuanced relationship between the number of retrieved texts and various performance metrics of a predicted causal network. Specifically, retrieving fewer texts generally enhances accuracy and precision. This could be attributed to a more focused selection of texts that potentially reduces noise and extraneous information, leading to sharper, more targeted predictions. On the other hand, increasing the number of retrieved texts tends to enhance the F1-score and AUROC. The broader text retrieval likely captures a wider array of true positives, which boosts the F1-score by balancing the precision and recall. Additionally, the increased context may enhance the model’s ability to discriminate between classes, reflected in higher AUROC values. 

\subsection{Faithfulness and scientificness of CoT reasoning (RQ3)}
This experiment studies whether the reasoning of CoT is faithful and scientific. We report results using papers collected every 5 years as well as using all 200 papers in RAG. Results in Figure \ref{main} show that the CoT reasoning of larger models tends to be less faithful and scientific, with a lower entailment rate and human evaluation score. This can be explained by the fact that larger models interpolate across vast amounts of data, often prioritizing fluency over fidelity. Their training objectives typically emphasize helpfulness and coherence over strict factual accuracy. While this makes them effective in open-ended tasks, it can also lead to reduced factuality. From Figure \ref{main}, we observe that models with higher entailment rates and greater faithfulness tend to achieve better F1-score. For example, LLaMA-3-8B, which exhibits the highest entailment rate, outperforms the other three models in causal network construction. This suggests that better reasoning based on retrieved text leads to a more accurate understanding of causal relationships. However, models with lower entailment rates can still achieve comparable performance—for instance, e.g. LLaMA-70B and Mixtral-8x7B-v0.1. This indicates that LLMs may still predict the correct causal relations even when their reasoning does not align closely with the retrieved content. This discrepancy may stem from the fact that larger models rely more heavily on internal knowledge or that their predictions diverge from explicit reasoning.

\begin{figure}[t!]
    \centering
    \includegraphics[width=1\linewidth]{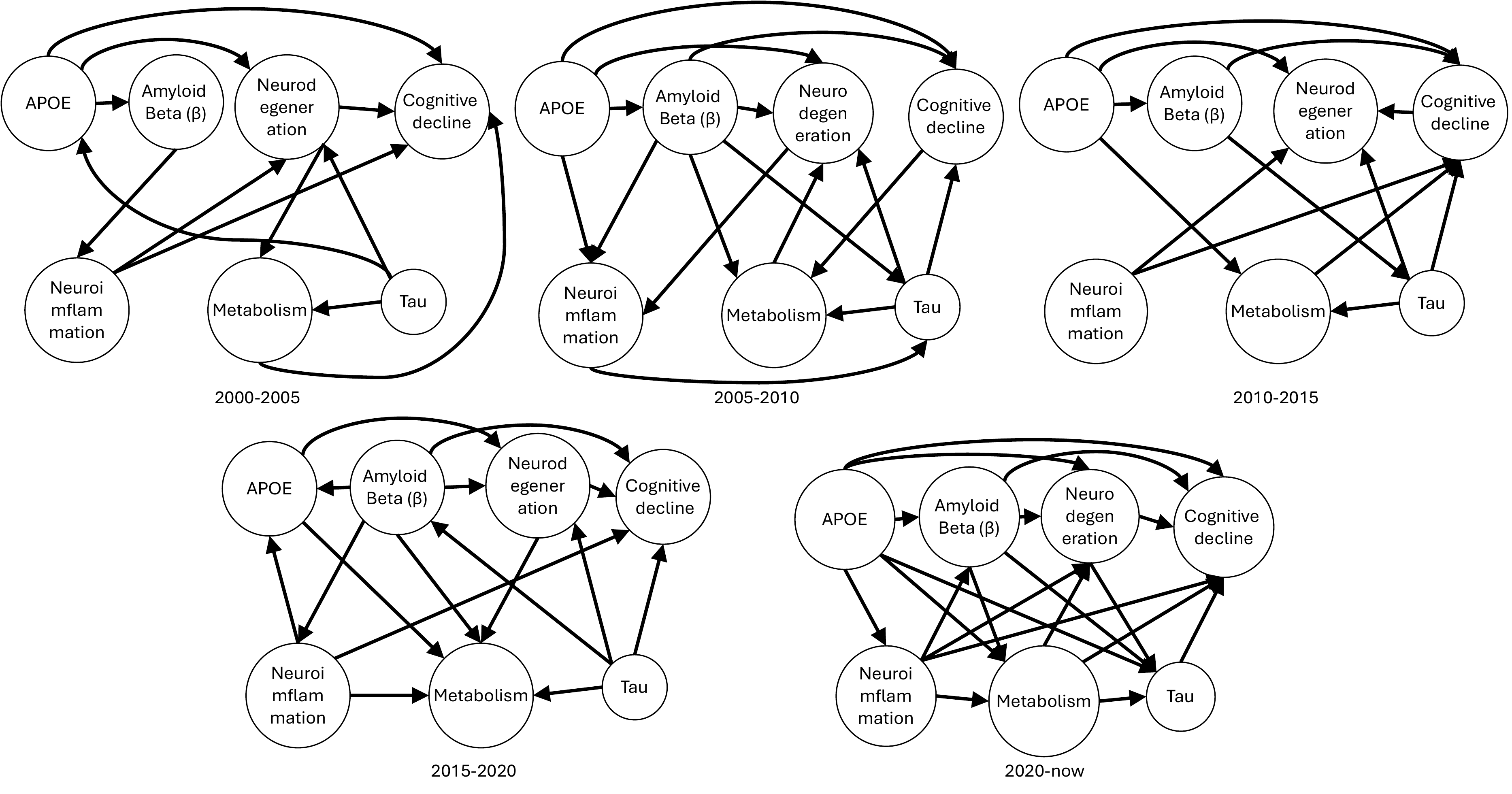}
    \caption{AD causal networks evolution in 25 years with a 5-year interval.}
    \label{fig:enter-label}
\end{figure}
\subsection{RAG for evolutionary causal networks of AD biomarkers (RQ4)}


In Figure \ref{fig:enter-label}, we show the evolution of the AD biomarkers causal network over the last 25 years in 5-year intervals using Meta-Llama-3-8B as the backbone LLM and $k$ is set to 50. Each figure exhibits variations in causal edges, which may be due to two factors:  1. The continuous advancement of human knowledge regarding AD biomarkers.  2. The inherent uncertainty in the model generation process.  

Overall, the causal network of AD biomarkers evolves, becoming increasingly complex as research uncovers more interactions among key factors. From \textbf{2000–2005}, the network is sparse, with only a few well-established causal links, primarily focusing on APOE, Amyloid Beta ($\beta$), and their direct effects. Additional connections emerge by \textbf{2005–2010}, incorporating neuroinflammation, metabolism, and tau, reflecting a growing understanding of the disease's multifactorial nature. In \textbf{2010–2015}, the network further expands, marking a shift toward recognizing multi-pathway interactions, where multiple upstream factors influence cognitive decline. After \textbf{2015}, the models become highly interconnected, integrating feedback loops and multi-scale influences. These advanced causal networks incorporate biomarker data, multi-omics insights, and AI-driven modeling, marking a transition from simple pathway analysis to sophisticated computational frameworks.  
\section{Conclusion}

This study introduces the novel application of RAG to integrate data from AD scientific literature and domain expertise, accelerating causal discovery in AD biomarkers that reflect the disease's pathophysiology. We present both standard and modified split-RAG models, incorporating uncertainty estimation to evaluate the reliability of the predicted causal connections by RAG. Our evaluation also involves both automated methods and human review to verify the LLM's reasoning for faithfulness and scientific accuracy. Our results show that RAG can improve LLMs' ability to construct causal networks, though LLMs generally remain limited in identifying causal relationships.

Looking ahead, we plan to compare different RAG models to determine their impact on the quality of causal networks. We also aim to investigate the low entailment rates observed, examining the performance of various entailment evaluation models and exploring why larger models like LLaMA-70B tend to produce lower entailment rates and less accurate networks. Further, we will explore extending our RAG framework to other scientific fields.



\bibliography{colm2025_conference}
\bibliographystyle{colm2025_conference}
\appendix

\end{document}